# Semiquantum secret sharing by using χ-type states

Ying Chen, Tian-Yu Ye*

College of Information & Electronic Engineering, Zhejiang Gongshang University, Hangzhou 310018, P.R.China

E-mail：happyyty@aliyun.com (T.Y.Ye)

**Abstract:** In this paper, a semiquantum secret sharing (SQSS) protocol based on χ-type states is proposed, which can accomplish the goal that only when two classical communicants cooperate together can they extract the shared secret key of a quantum communicant. Detailed security analysis turns out that this protocol is completely robust against an eavesdropper. This protocol has some merits: (1) it only requires one kind of quantum entangled state as the initial quantum resource; (2) it doesn't employ quantum entanglement swapping or unitary operations; and (3) it needn't share private keys among different participants beforehand.

## 1 Introduction

Quantum cryptography, as an important branch of quantum information, combines the laws of quantum mechanics and the characteristics of classical cryptography to accomplish the secret communication of information via the quantum channel. Currently, quantum cryptography has gained numerous branches, such as quantum key distribution (QKD) [1-5], quantum secure direct communication (QSDC) [6-13], quantum dialogue (QD) [14-20], quantum secret sharing (QSS) [21-30], *etc*. Here, the basic principle of QSS is: the sender splits the secret message into several pieces, and distributes them to the receivers, respectively; and none of the receivers can reveal the secret message without the collaboration of other receivers. The first QSS protocol was put forward by Hillery *et al.* [21] in 1999 based on the three-particle and four-particle GHZ entangled states. Since then, many QSS protocols have been proposed based on various quantum entangled states, such as Bell states [22-25], GHZ states [26-29], W states [30] and so on. However, the previous QSS protocols [21-30] always require all communicants to have complete quantum capabilities, which may be impractical in some circumstances, as partial of them may not have the ability to afford expensive quantum resources and operations.

In 2007, Boyer *et al.* [31,32] proposed two earliest semiquantum key distribution (SQKD) protocols with the measure-resend characteristic and the randomization characteristic, respectively, each of which only requires one party to have full quantum capabilities, and enables key establishment between a quantum user and a classical one. A classical user is widely considered to be limited within the following operations [31,32]: (a) transmitting particles via the quantum channel; (b) measuring particles in the $Z$ basis (i.e., $\{|0\rangle, |1\rangle\}$); (c) preparing the fresh particles in the $Z$ basis; and (d) reordering particles via different delay lines. Later, in 2009, Zou *et al.* [33] were devoted to reducing the number of kinds of initial quantum states for SQKD, and especially put forward the first single-state SQKD; in 2015, Krawec [34] successfully designed a two-user

SQKD protocol by introducing a quantum third party to help establish a private key between two classical users; in 2016, Krawec [35] successfully made the reflection operations of the classical user contribute to the establishment of a secret key; in 2020 and 2022, Ye *et al.* [36,37] utilized single photons in two degrees of freedom to construct SQKD protocols; authenticated SQKD protocols without an authenticated channel [38,39] have also been designed; Refs.[40,41] concentrated on designing high-dimensional SQKD; different quantum entangled states have also been used to design SQKD protocols [34,38,39,41-44].

Besides the semiquantum protocol construction techniques, scholars have also studied how to evaluate the security of semiquantum protocols [45]. Boyer *et al.* first gave the definition of robustness in Refs.[31,32], which can be classified into three types: complete robustness, complete nonrobustness and partial robustness. Another security analysis method is the information-theoretic analysis, which aims to know exactly how much information an eavesdropper can gain given a certain amount of detectable noise. Three approaches for calculating the information-theoretic bound on the key rate of SQKD protocols have been put forward [46-49].

Besides SQKD, the novel branches of semiquantum secure direct communication (QSDC) [50-56], semiquantum dialogue (QD) [57-60], semiquantum private comparison (SQPC) [61-66], semiquantum secret sharing (QSS) [67-74], *etc*, have also gained considerable developments. The basic principle of SQSS generally is: the quantum sender splits the secret message into several pieces, and transmits them to the classical receivers, respectively; and only when all classical receivers cooperate together can they recover the quantum sender's secret message. In 2010, Li *et al.* [67] proposed two SQSS protocols by using GHZ-like states; in 2011,Wang *et al.* [68] suggested an SQSS protocol based on two-particle entangled states; in 2013, Li *et al.* [69] utilized product states to put forward an SQSS protocol, and Yang and Hwang [70] proposed an improvement for this protocol; in 2016, Gao *et al.* [71] proposed an SQSS scheme based on rearranging orders of qubits; in 2018,Ye *et al.* [72] put forward two circular SQSS protocols by using single particles, where the particles prepared by the quantum party are transmitted in a circular way; in 2019, Tsai *et al.* [73] constructed an SQSS protocol by using W states; and in 2021, Li *et al.* [74] constructed an SQSS protocol with four-particle Cluster states and Bell states by using quantum entanglement swapping.

However, the SQSS protocol in Ref.[74] needs two kinds of quantum entangled states as the initial quantum resource and quantum entanglement swapping. At present, χ-type states have never been adopted to design the SQSS protocol. In order to ease the burdens on the preparation of initial quantum resource and the usage of quantum entanglement swapping in the SQSS protocol based on four-particle quantum entangled states, in this paper, we are devoted to utilizing the entanglement correlation property of χ-type states to design a novel SQSS protocol also with four-particle quantum entangled states. We also validate the complete robustness of the proposed SQSS protocol in detail, which means that nonzero information obtained by any eavesdropper on the shared private key implies detectable errors on the tested bits [31,32]. The proposed SQSS protocol has the following merits: (1) it only requires one kind of quantum entangled state as the initial quantum resource; (2) it doesn't employ quantum entanglement swapping or unitary operations; and (3) it needn't share private keys among different participants beforehand.

## 2  Protocol description

In this section, we describe the proposed three-party SQSS protocol with χ-type states. Define one χ-type state as

$$|\chi^{00}\rangle_{1234} = \frac{\sqrt{2}}{4}(|0000\rangle - |0101\rangle + |0011\rangle + |0110\rangle + |1001\rangle + |1010\rangle + |1100\rangle - |1111\rangle)_{1234} \tag{1}$$

$$= \frac{1}{2}(|\phi^+\rangle|00\rangle + |\phi^-\rangle|11\rangle - |\psi^-\rangle|01\rangle + |\psi^+\rangle|10\rangle)_{1234} \tag{2}$$

$$= \frac{1}{2}(|00\rangle|\phi^+\rangle + |11\rangle|\phi^-\rangle - |01\rangle|\psi^-\rangle + |10\rangle|\psi^+\rangle)_{1234}, \tag{3}$$

$$= \frac{1}{2}(|0\rangle|\phi^+\rangle|0\rangle + |0\rangle|\psi^-\rangle|1\rangle + |1\rangle|\psi^+\rangle|0\rangle + |1\rangle|\phi^-\rangle|1\rangle)_{1234}, \tag{4}$$

where $|\phi^\pm\rangle = \frac{1}{\sqrt{2}}(|00\rangle \pm |11\rangle)$ and $|\psi^\pm\rangle = \frac{1}{\sqrt{2}}(|01\rangle \pm |10\rangle)$. An orthonormal basis set in four-qubit Hilbert space can be derived from imposing Pauli operations on qubits 1 and 3 of $|\chi^{00}\rangle_{1234}$:

$$FMB = \left\{ |\chi^{kl}\rangle_{1234} = \sigma_1^k \sigma_3^l |\chi^{00}\rangle_{1234} \mid k,l = 0,1,2,3 \right\}, \tag{5}$$

where $\sigma^0 = |0\rangle\langle 0| + |1\rangle\langle 1|$, $\sigma^1 = |0\rangle\langle 1| + |1\rangle\langle 0|$, $\sigma^2 = |0\rangle\langle 1| - |1\rangle\langle 0|$ and $\sigma^3 = |0\rangle\langle 0| - |1\rangle\langle 1|$ are four Pauli operations.

Assume that Alice is the party with complete quantum capabilities, while Bob and Charlie are two parties only equipped with limited quantum capabilities. Moreover, Alice wants to share a secret key with Bob and Charlie on the condition that none of Bob and Charlie can recover this secret key alone. We put forward the following SQSS protocol to accomplish this goal. Here, $|0\rangle$ and $|1\rangle$ correspond to the classical bits 0 and 1, respectively.

Step 1: Alice prepares $8n$ χ-type states all in the state of $|\chi^{00}\rangle$ and divides them into four different particle sequences $S_1, S_2, S_3$ and $S_4$. Here, $S_j$ is the sequence composed by all of the $j$ th particles from these $8n$ χ-type states, where $j = 1,2,3,4$. Alice retains $S_2$ and $S_3$ in her hand, and sends the particles of $S_1$ ($S_4$) to Bob (Charlie) one by one via the quantum channel. Note that except the first particle of $S_1$ ($S_4$), Alice sends out the next one only after receiving the previous one.

Step 2: Upon receiving each particle of $S_1$, Bob randomly chooses either to measure the qubit with the $Z$ basis, prepare a fresh one in the found state and send it to Alice (referred as SIFT), or to reflect it back to Alice without disturbance (referred as CTRL).

Upon receiving each particle of $S_4$, Charlie also randomly chooses either to SIFT or to CTRL.

Step 3: Alice temporarily stores all of the particles from Bob and Charlie. Bob and Charlie announce the positions where they chose to SIFT.

Step 4: According to Bob and Charlie's choices, Alice performs one of the four operations, as illustrated in Table 1. Here, $S_1^i$, $S_2^i$, $S_3^i$ and $S_4^i$ represent the $i$ th particles of $S_1$, $S_2$, $S_3$ and $S_4$, respectively, where $i \in \{1, 2, \ldots, 8n\}$.

Table 1   Three parties' operations on the particles

| Case | Bob's operation | Charlie's operation | Alice's operation |
|---|---|---|---|
| (a) | SIFT | SIFT | ACTION 1 |
| (b) | SIFT | CTRL | ACTION 2 |
| (c) | CTRL | SIFT | ACTION 3 |
| (d) | CTRL | CTRL | ACTION 4 |

ACTION 1: To perform Bell basis measurement on $S_2^i$ and $S_3^i$, and measure the fresh particle corresponding to $S_1^i$ from Bob and the fresh particle corresponding to $S_4^i$ from Charlie with the $Z$ basis, respectively;

ACTION 2: To perform Bell basis measurement on $S_3^i$ and $S_4^i$, and measure the fresh particle corresponding to $S_1^i$ from Bob and $S_2^i$ with the $Z$ basis, respectively;

ACTION 3: To perform Bell basis measurement on $S_1^i$ and $S_2^i$, and measure $S_3^i$ and the fresh particle corresponding to $S_4^i$ from Charlie with the $Z$ basis, respectively;

ACTION 4: To measure $S_1^i$, $S_2^i$, $S_3^i$ and $S_4^i$ with the *FMB* basis.

Case (a): When both Bob and Charlie have chosen to SIFT, Alice implements ACTION 1. Note that this Case is for both security check and secret sharing. If no eavesdropper exists, Alice's Bell basis measurement result on $S_2^i$ and $S_3^i$, Bob's $Z$ basis measurement result on $S_1^i$ and Charlie's $Z$ basis measurement result on $S_4^i$ should obey to Eq.(4); Alice's $Z$ basis measurement result on the fresh particle corresponding to $S_1^i$ ($S_4^i$) from Bob (Charlie) should be same to the prepared state of the fresh particle corresponding to $S_1^i$ ($S_4^i$) from Bob (Charlie). To check the existence of Eve, Alice, Bob and Charlie randomly choose $n$ positions of this Case; Bob and Charlie need to tell Alice their measurement results on these $n$ chosen positions;

Case (b): When Bob has chosen to SIFT and Charlie has chosen to CTRL, Alice performs ACTION 2 to check whether there is an eavesdropper or not. If no eavesdropper exists, Alice's Bell basis measurement result on $S_3^i$ and $S_4^i$, Bob's $Z$ basis measurement result on $S_1^i$ and Alice's $Z$ basis measurement result on $S_2^i$ should obey to Eq.(3); and Alice's $Z$ basis measurement result on the fresh particle corresponding to $S_1^i$ from Bob should be same to the prepared state of the fresh particle corresponding to $S_1^i$ from Bob. To check the existence of Eve, Bob needs to tell Alice his measurement result on $S_1^i$ in this Case;

Case (c): When Charlie has chosen to SIFT and Bob has chosen to CTRL, Alice performs ACTION 3 to check whether there is an eavesdropper or not. If no eavesdropper exists, Alice's Bell basis measurement result on $S_1^i$ and $S_2^i$, Alice's $Z$ basis measurement result on $S_3^i$ and Charlie's $Z$ basis measurement result on $S_4^i$ should obey to Eq.(2); and Alice's $Z$ basis measurement result on the fresh particle corresponding to $S_4^i$ from Charlie should be same to the prepared state of the fresh particle corresponding to $S_4^i$ from Charlie. To check the existence of Eve, Charlie needs to tell Alice her measurement result on $S_4^i$ in this Case;

Case (d): When both Bob and Charlie have chosen to CTRL, Alice performs ACTION 4 to check whether there is an eavesdropper or not. If no eavesdropper exists, Alice's *FMB* basis measurement result on $S_1^i$, $S_2^i$, $S_3^i$ and $S_4^i$ should always be $\left|\chi^{00}\right\rangle$, according to Eq.(1).

Step 5: Alice checks the error rates in Cases (a), (b), (c) and (d). If the error rate in any of these four Cases is abnormally high, the protocol will be aborted immediately; otherwise, it will be carried on.

Step 6: The remaining $n$ positions in Case (a) are used for secret sharing. Denote $r_b^k$ ($r_c^k$) as the classical bit corresponding to Bob's (Charlie's) $Z$ basis measurement result on the $k$ th remaining position of Case (a), where $k = 1, 2, \ldots, n$. Since Alice has used the $Z$ basis to measure the corresponding fresh particle from Bob (Charlie) on the $k$ th remaining position of Case (a), Alice can automatically know $r_b^k$ ($r_c^k$). In this way, Alice encodes the corresponding shared secret key bit as $r_a^k = r_b^k \oplus r_c^k$. Apparently, Bob and Charlie can recover $r_a^k$ only when they collaborate. For clarity, the relations among the classical bits related to three parties' $Z$ basis measurement results on the remaining position of Case (a) and the corresponding shared secret key bit are shown in Table 2.

Table 2  The classical bits related to three parties' Z basis measurement results on the remaining position of Case (a) and the corresponding shared secret key bit

| The classical bit related to Bob's measurement result | The classical bit related to Charlie's measurement result | The classical bits related to Alice's measurement results | Shared secret key bit |
| --- | --- | --- | --- |
| 0 | 0 | 00 | 0 |
| 0 | 1 | 01 | 1 |
| 1 | 0 | 10 | 1 |
| 1 | 1 | 11 | 0 |

## 3  Security analysis

### 3.1  The outside attack

In the following, we validate the complete robustness of the proposed SQSS protocol against an outside eavesdropper.

An outside eavesdropper, Eve, may launch the entangle-measure attack to obtain something useful about Alice's shared secret key bits. Eve's entangle-measure attack can be depicted as Fig.1, which is comprised of two unitaries, $U_E$ and $U_F$. Here, $U_E$ attacks particles going from Alice to Bob and Charlie, while $U_F$ attacks particles going back from Bob and Charlie to Alice; $U_E$ and $U_F$ share a common probe space with initial state $|\xi\rangle_E$. As illustrated in Refs.[31,32], the shared probe permits Eve to launch the attack on the returned particles by depending on the knowledge gained from $U_E$; and any attack where Eve would let $U_F$ depend on a measurement after $U_E$ can be accomplished by $U_E$ and $U_F$ with controlled gates. In the following, we will validate Theorem 1 in detail.

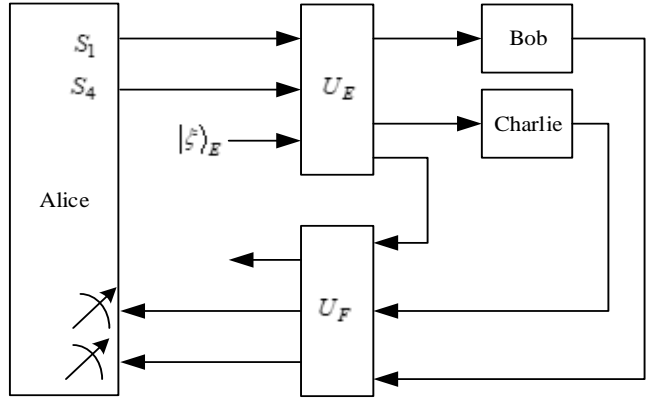

Fig.1  Eve's entangle-measure attack with two unitaries $U_E$ and $U_F$

**Theorem 1:** When Eve performs attack $(U_E, U_F)$ on the particles from Alice to Bob and Charlie and back to Alice, for incurring no error in Step 4, the final state of Eve's probe should be independent from Alice, Bob and Charlie's operations and measurement results. Therefore, Eve gets no information on Alice's shared secret key bit.

**Proof:** The effect of $U_E$ on the qubits $|0\rangle$ and $|1\rangle$ can be depicted as

$$U_E(|0\rangle|\xi\rangle_E) = \beta_{00}|0\rangle|\xi_{00}\rangle + \beta_{01}|1\rangle|\xi_{01}\rangle, \quad (6)$$

$$U_E(|1\rangle|\xi\rangle_E) = \beta_{10}|0\rangle|\xi_{10}\rangle + \beta_{11}|1\rangle|\xi_{11}\rangle, \quad (7)$$

where $|\xi_{00}\rangle, |\xi_{01}\rangle, |\xi_{10}\rangle$ and $|\xi_{11}\rangle$ are Eve's probe states determined by $U_E$, $|\beta_{00}|^2 + |\beta_{01}|^2 = 1$ and $|\beta_{10}|^2 + |\beta_{11}|^2 = 1$.

According to Stinespring dilation theorem, the global state of the composite system before Bob and Charlie's operation is

$$U_E(|\chi^{00}\rangle_{1234} \otimes |\xi\rangle_E) = U_E\left[\frac{\sqrt{2}}{4}(|0000\rangle - |0101\rangle + |0011\rangle + |0110\rangle + |1001\rangle + |1010\rangle + |1100\rangle - |1111\rangle)_{1234} \otimes |\xi\rangle_E\right]$$

$$= \frac{\sqrt{2}}{4}\Big[(\beta_{00}|0\rangle_1|\xi_{00}\rangle + \beta_{01}|1\rangle_1|\xi_{01}\rangle)|00\rangle_{23}(\beta_{00}|0\rangle_4|\xi_{00}\rangle + \beta_{01}|1\rangle_4|\xi_{01}\rangle)$$

$$-(\beta_{00}|0\rangle_1|\xi_{00}\rangle + \beta_{01}|1\rangle_1|\xi_{01}\rangle)|10\rangle_{23}(\beta_{10}|0\rangle_4|\xi_{10}\rangle + \beta_{11}|1\rangle_4|\xi_{11}\rangle)$$

$$+(\beta_{00}|0\rangle_1|\xi_{00}\rangle + \beta_{01}|1\rangle_1|\xi_{01}\rangle)|01\rangle_{23}(\beta_{10}|0\rangle_4|\xi_{10}\rangle + \beta_{11}|1\rangle_4|\xi_{11}\rangle)$$

$$+(\beta_{00}|0\rangle_1|\xi_{00}\rangle + \beta_{01}|1\rangle_1|\xi_{01}\rangle)|11\rangle_{23}(\beta_{00}|0\rangle_4|\xi_{00}\rangle + \beta_{01}|1\rangle_4|\xi_{01}\rangle)$$

$$+(\beta_{10}|0\rangle_1|\xi_{10}\rangle + \beta_{11}|1\rangle_1|\xi_{11}\rangle)|00\rangle_{23}(\beta_{10}|0\rangle_4|\xi_{10}\rangle + \beta_{11}|1\rangle_4|\xi_{11}\rangle)$$

$$+(\beta_{10}|0\rangle_1|\xi_{10}\rangle + \beta_{11}|1\rangle_1|\xi_{11}\rangle)|01\rangle_{23}(\beta_{00}|0\rangle_4|\xi_{00}\rangle + \beta_{01}|1\rangle_4|\xi_{01}\rangle)$$

$$+(\beta_{10}|0\rangle_1|\xi_{10}\rangle + \beta_{11}|1\rangle_1|\xi_{11}\rangle)|10\rangle_{23}(\beta_{00}|0\rangle_4|\xi_{00}\rangle + \beta_{01}|1\rangle_4|\xi_{01}\rangle)$$

$$-(\beta_{10}|0\rangle_1|\xi_{10}\rangle + \beta_{11}|1\rangle_1|\xi_{11}\rangle)|11\rangle_{23}(\beta_{10}|0\rangle_4|\xi_{10}\rangle + \beta_{11}|1\rangle_4|\xi_{11}\rangle)\Big]$$

$$= \frac{\sqrt{2}}{4}\Big[|0\rangle_1|00\rangle_{23}|0\rangle_4(\beta_{00}^2|\xi_{00}\rangle|\xi_{00}\rangle + \beta_{10}^2|\xi_{10}\rangle|\xi_{10}\rangle) + |0\rangle_1|00\rangle_{23}|1\rangle_4(\beta_{00}\beta_{01}|\xi_{00}\rangle|\xi_{01}\rangle + \beta_{10}\beta_{11}|\xi_{10}\rangle|\xi_{11}\rangle)$$

$$+ |1\rangle_1|00\rangle_{23}|0\rangle_4(\beta_{01}\beta_{00}|\xi_{01}\rangle|\xi_{00}\rangle + \beta_{11}\beta_{10}|\xi_{11}\rangle|\xi_{10}\rangle) + |1\rangle_1|00\rangle_{23}|1\rangle_4(\beta_{01}^2|\xi_{01}\rangle|\xi_{01}\rangle + \beta_{11}^2|\xi_{11}\rangle|\xi_{11}\rangle)$$

$$+ |0\rangle_1|10\rangle_{23}|0\rangle_4(\beta_{10}\beta_{00}|\xi_{10}\rangle|\xi_{00}\rangle - \beta_{00}\beta_{10}|\xi_{00}\rangle|\xi_{10}\rangle) + |0\rangle_1|10\rangle_{23}|1\rangle_4(\beta_{10}\beta_{01}|\xi_{10}\rangle|\xi_{01}\rangle - \beta_{00}\beta_{11}|\xi_{00}\rangle|\xi_{11}\rangle)$$

$$+ |1\rangle_1|10\rangle_{23}|0\rangle_4(\beta_{11}\beta_{00}|\xi_{11}\rangle|\xi_{00}\rangle - \beta_{01}\beta_{10}|\xi_{01}\rangle|\xi_{10}\rangle) + |1\rangle_1|10\rangle_{23}|1\rangle_4(\beta_{11}\beta_{01}|\xi_{11}\rangle|\xi_{01}\rangle - \beta_{01}\beta_{11}|\xi_{01}\rangle|\xi_{11}\rangle)$$

$$+ |0\rangle_1|01\rangle_{23}|0\rangle_4(\beta_{00}\beta_{10}|\xi_{00}\rangle|\xi_{10}\rangle + \beta_{10}\beta_{00}|\xi_{10}\rangle|\xi_{00}\rangle) + |0\rangle_1|01\rangle_{23}|1\rangle_4(\beta_{00}\beta_{11}|\xi_{00}\rangle|\xi_{11}\rangle + \beta_{10}\beta_{01}|\xi_{10}\rangle|\xi_{01}\rangle)$$

$$+ |1\rangle_1|01\rangle_{23}|0\rangle_4(\beta_{01}\beta_{10}|\xi_{01}\rangle|\xi_{10}\rangle + \beta_{11}\beta_{00}|\xi_{11}\rangle|\xi_{00}\rangle) + |1\rangle_1|01\rangle_{23}|1\rangle_4(\beta_{01}\beta_{11}|\xi_{01}\rangle|\xi_{11}\rangle + \beta_{11}\beta_{01}|\xi_{11}\rangle|\xi_{01}\rangle)$$

$$+ |0\rangle_1|11\rangle_{23}|0\rangle_4(\beta_{00}^2|\xi_{00}\rangle|\xi_{00}\rangle - \beta_{10}^2|\xi_{10}\rangle|\xi_{10}\rangle) + |0\rangle_1|11\rangle_{23}|1\rangle_4(\beta_{00}\beta_{01}|\xi_{00}\rangle|\xi_{01}\rangle - \beta_{10}\beta_{11}|\xi_{10}\rangle|\xi_{11}\rangle)$$

$$+ |1\rangle_1|11\rangle_{23}|0\rangle_4(\beta_{01}\beta_{00}|\xi_{01}\rangle|\xi_{00}\rangle - \beta_{11}\beta_{10}|\xi_{11}\rangle|\xi_{10}\rangle) + |1\rangle_1|11\rangle_{23}|1\rangle_4(\beta_{01}^2|\xi_{01}\rangle|\xi_{01}\rangle - \beta_{11}^2|\xi_{11}\rangle|\xi_{11}\rangle)\Big]$$

$$= \frac{\sqrt{2}}{4}\Big(|0\rangle_1|00\rangle_{23}|0\rangle_4|E_{0000}\rangle + |0\rangle_1|00\rangle_{23}|1\rangle_4|E_{0001}\rangle + |1\rangle_1|00\rangle_{23}|0\rangle_4|E_{1000}\rangle + |1\rangle_1|00\rangle_{23}|1\rangle_4|E_{1001}\rangle$$

$$+ |0\rangle_1|10\rangle_{23}|0\rangle_4|E_{0100}\rangle + |0\rangle_1|10\rangle_{23}|1\rangle_4|E_{0101}\rangle + |1\rangle_1|10\rangle_{23}|0\rangle_4|E_{1100}\rangle + |1\rangle_1|10\rangle_{23}|1\rangle_4|E_{1101}\rangle$$

$$+ |0\rangle_1|01\rangle_{23}|0\rangle_4|E_{0010}\rangle + |0\rangle_1|01\rangle_{23}|1\rangle_4|E_{0011}\rangle + |1\rangle_1|01\rangle_{23}|0\rangle_4|E_{1010}\rangle + |1\rangle_1|01\rangle_{23}|1\rangle_4|E_{1011}\rangle$$

$$+|0\rangle_1|11\rangle_{23}|0\rangle_4|E_{0110}\rangle+|0\rangle_1|11\rangle_{23}|1\rangle_4|E_{0111}\rangle+|1\rangle_1|11\rangle_{23}|0\rangle_4|E_{1110}\rangle+|1\rangle_1|11\rangle_{23}|1\rangle_4|E_{1111}\rangle), \quad (8)$$

where the subscripts 1, 2, 3 and 4 represent the particles from $S_1, S_2, S_3$ and $S_4$, respectively. Besides, $|E_{0000}\rangle = \beta_{00}^2|\xi_{00}\rangle|\xi_{00}\rangle + \beta_{10}^2|\xi_{10}\rangle|\xi_{10}\rangle$, $|E_{0001}\rangle = \beta_{00}\beta_{01}|\xi_{00}\rangle|\xi_{01}\rangle + \beta_{10}\beta_{11}|\xi_{10}\rangle|\xi_{11}\rangle$,

$|E_{1000}\rangle = \beta_{01}\beta_{00}|\xi_{01}\rangle|\xi_{00}\rangle + \beta_{11}\beta_{10}|\xi_{11}\rangle|\xi_{10}\rangle$, $|E_{1001}\rangle = \beta_{01}^2|\xi_{01}\rangle|\xi_{01}\rangle + \beta_{11}^2|\xi_{11}\rangle|\xi_{11}\rangle$,

$|E_{0100}\rangle = \beta_{10}\beta_{00}|\xi_{10}\rangle|\xi_{00}\rangle - \beta_{00}\beta_{10}|\xi_{00}\rangle|\xi_{10}\rangle$, $|E_{0101}\rangle = \beta_{10}\beta_{01}|\xi_{10}\rangle|\xi_{01}\rangle - \beta_{00}\beta_{11}|\xi_{00}\rangle|\xi_{11}\rangle$,

$|E_{1100}\rangle = \beta_{11}\beta_{00}|\xi_{11}\rangle|\xi_{00}\rangle - \beta_{01}\beta_{10}|\xi_{01}\rangle|\xi_{10}\rangle$, $|E_{1101}\rangle = \beta_{11}\beta_{01}|\xi_{11}\rangle|\xi_{01}\rangle - \beta_{01}\beta_{11}|\xi_{01}\rangle|\xi_{11}\rangle$,

$|E_{0010}\rangle = \beta_{00}\beta_{10}|\xi_{00}\rangle|\xi_{10}\rangle + \beta_{10}\beta_{00}|\xi_{10}\rangle|\xi_{00}\rangle$, $|E_{0011}\rangle = \beta_{00}\beta_{11}|\xi_{00}\rangle|\xi_{11}\rangle + \beta_{10}\beta_{01}|\xi_{10}\rangle|\xi_{01}\rangle$,

$|E_{1010}\rangle = \beta_{01}\beta_{10}|\xi_{01}\rangle|\xi_{10}\rangle + \beta_{11}\beta_{00}|\xi_{11}\rangle|\xi_{00}\rangle$, $|E_{1011}\rangle = \beta_{01}\beta_{11}|\xi_{01}\rangle|\xi_{11}\rangle + \beta_{11}\beta_{01}|\xi_{11}\rangle|\xi_{01}\rangle$,

$|E_{0110}\rangle = \beta_{00}^2|\xi_{00}\rangle|\xi_{00}\rangle - \beta_{10}^2|\xi_{10}\rangle|\xi_{10}\rangle$, $|E_{0111}\rangle = \beta_{00}\beta_{01}|\xi_{00}\rangle|\xi_{01}\rangle - \beta_{10}\beta_{11}|\xi_{10}\rangle|\xi_{11}\rangle$,

$|E_{1110}\rangle = \beta_{01}\beta_{00}|\xi_{01}\rangle|\xi_{00}\rangle - \beta_{11}\beta_{10}|\xi_{11}\rangle|\xi_{10}\rangle$, $|E_{1111}\rangle = \beta_{01}^2|\xi_{01}\rangle|\xi_{01}\rangle - \beta_{11}^2|\xi_{11}\rangle|\xi_{11}\rangle$.

(i) Assume that both Bob and Charlie have chosen to SIFT. In order that Eve's attacks will not be discovered in Case (a) of Step 4, Alice's Bell basis measurement result on $S_2^i$ and $S_3^i$, Bob's Z basis measurement result on $S_1^i$ and Charlie's Z basis measurement result on $S_4^i$ should obey to Eq.(4). Thus, it should satisfy that

$$|E_{0100}\rangle = |E_{0010}\rangle = |E_{0001}\rangle = |E_{0111}\rangle = |E_{1000}\rangle = |E_{1110}\rangle = |E_{1101}\rangle = |E_{1011}\rangle = 0, \quad (9)$$

$$|E_{0000}\rangle = |E_{0110}\rangle, \quad (10)$$

$$|E_{0101}\rangle = -|E_{0011}\rangle, \quad (11)$$

$$|E_{1100}\rangle = |E_{1010}\rangle, \quad (12)$$

$$|E_{1001}\rangle = -|E_{1111}\rangle. \quad (13)$$

Inserting Eqs.(9-13) into Eq.(8) generates

$$U_E(|\chi^{00}\rangle_{1234} \otimes |\xi\rangle_E) = \frac{1}{2}(|0\rangle_1|\phi^+\rangle_{23}|0\rangle_4|E_{0000}\rangle + |1\rangle_1|\phi^-\rangle_{23}|1\rangle_4|E_{1001}\rangle$$

$$+|0\rangle_1|\psi^-\rangle_{23}|1\rangle_4|E_{0011}\rangle + |1\rangle_1|\psi^+\rangle_{23}|0\rangle_4|E_{1100}\rangle). \quad (14)$$

Let $|\vartheta_{00}\rangle = |\phi^+\rangle$, $|\vartheta_{01}\rangle = |\psi^-\rangle$, $|\vartheta_{10}\rangle = |\psi^+\rangle$, $|\vartheta_{11}\rangle = |\phi^-\rangle$, $|\Delta_{00}\rangle = |E_{0000}\rangle$, $|\Delta_{01}\rangle = |E_{0011}\rangle$, $|\Delta_{10}\rangle = |E_{1100}\rangle$, $|\Delta_{11}\rangle = |E_{1001}\rangle$. Then, Eq.(14) becomes

$$U_E(|\chi^{00}\rangle_{1234} \otimes |\xi\rangle_E) = \frac{1}{2}(|0\rangle_1|\vartheta_{00}\rangle_{23}|0\rangle_4|\Delta_{00}\rangle + |1\rangle_1|\vartheta_{11}\rangle_{23}|1\rangle_4|\Delta_{11}\rangle$$

$$+|0\rangle_1|\vartheta_{01}\rangle_{23}|1\rangle_4|\Delta_{01}\rangle + |1\rangle_1|\vartheta_{10}\rangle_{23}|0\rangle_4|\Delta_{10}\rangle). \quad (15)$$

Moreover, Alice's $Z$ basis measurement result on the fresh particle corresponding to $S_1^i$ ($S_4^i$) from Bob (Charlie) should be same to the prepared state from Bob (Charlie). As a result, $U_F$ should establish the following relations:

$$U_F\left(|x\rangle_1|\vartheta_{xy}\rangle_{23}|y\rangle_4|\Delta_{xy}\rangle\right) = |x\rangle_1|\vartheta_{xy}\rangle_{23}|y\rangle_4|\Omega_{xy}\rangle, \quad x, y \in \{0,1\}, \tag{16}$$

which implies that $U_F$ cannot change the states of the fresh particle corresponding to $S_1^i$ from Bob and the fresh particle corresponding to $S_4^i$ from Charlie after Bob and Charlie's operations. Otherwise, the probability of Eve's being detected will be non-zero.

(ii) Assume that Bob has chosen to SIFT and Charlie has chosen to CRTL. Thus, when Bob's measurement result is $|0\rangle$, the state of the composite system in Eq.(8) is evolved into

$|0\rangle_1|00\rangle_{23}|0\rangle_4|E_{0000}\rangle + |0\rangle_1|00\rangle_{23}|1\rangle_4|E_{0001}\rangle + |0\rangle_1|10\rangle_{23}|0\rangle_4|E_{0100}\rangle + |0\rangle_1|10\rangle_{23}|1\rangle_4|E_{0101}\rangle$
$+|0\rangle_1|01\rangle_{23}|0\rangle_4|E_{0010}\rangle + |0\rangle_1|01\rangle_{23}|1\rangle_4|E_{0011}\rangle + |0\rangle_1|11\rangle_{23}|0\rangle_4|E_{0110}\rangle + |0\rangle_1|11\rangle_{23}|1\rangle_4|E_{0111}\rangle$ ; and when Bob's measurement result is $|1\rangle$, the state of the composite system in Eq.(8) is evolved into

$|1\rangle_1|00\rangle_{23}|0\rangle_4|E_{1000}\rangle + |1\rangle_1|00\rangle_{23}|1\rangle_4|E_{1001}\rangle + |1\rangle_1|10\rangle_{23}|0\rangle_4|E_{1100}\rangle + |1\rangle_1|10\rangle_{23}|1\rangle_4|E_{1101}\rangle$
$+|1\rangle_1|01\rangle_{23}|0\rangle_4|E_{1010}\rangle + |1\rangle_1|01\rangle_{23}|1\rangle_4|E_{1011}\rangle + |1\rangle_1|11\rangle_{23}|0\rangle_4|E_{1110}\rangle + |1\rangle_1|11\rangle_{23}|1\rangle_4|E_{1111}\rangle$ .

Firstly, consider that Bob's measurement result is $|0\rangle$. After Eve performs $U_F$ on the particles from Bob and Charlie back to Alice, according to Eqs.(9-11,16), the state of the composite system is changed into

$U_F\left(|0\rangle_1|00\rangle_{23}|0\rangle_4|E_{0000}\rangle + |0\rangle_1|00\rangle_{23}|1\rangle_4|E_{0001}\rangle + |0\rangle_1|10\rangle_{23}|0\rangle_4|E_{0100}\rangle + |0\rangle_1|10\rangle_{23}|1\rangle_4|E_{0101}\rangle\right.$
$\left.+|0\rangle_1|01\rangle_{23}|0\rangle_4|E_{0010}\rangle + |0\rangle_1|01\rangle_{23}|1\rangle_4|E_{0011}\rangle + |0\rangle_1|11\rangle_{23}|0\rangle_4|E_{0110}\rangle + |0\rangle_1|11\rangle_{23}|1\rangle_4|E_{0111}\rangle\right)$

$= U_F\left(|0\rangle_1|00\rangle_{23}|0\rangle_4|E_{0000}\rangle + |0\rangle_1|10\rangle_{23}|1\rangle_4|E_{0101}\rangle + |0\rangle_1|01\rangle_{23}|1\rangle_4|E_{0011}\rangle + |0\rangle_1|11\rangle_{23}|0\rangle_4|E_{0110}\rangle\right)$

$= \sqrt{2}U_F\left(|0\rangle_1|\phi^+\rangle_{23}|0\rangle_4|E_{0000}\rangle + |0\rangle_1|\psi^-\rangle_{23}|1\rangle_4|E_{0011}\rangle\right)$

$= \sqrt{2}U_F\left(|0\rangle_1|\vartheta_{00}\rangle_{23}|0\rangle_4|\Delta_{00}\rangle + |0\rangle_1|\vartheta_{01}\rangle_{23}|1\rangle_4|\Delta_{01}\rangle\right)$

$= \sqrt{2}\left(|0\rangle_1|\vartheta_{00}\rangle_{23}|0\rangle_4|\Omega_{00}\rangle + |0\rangle_1|\vartheta_{01}\rangle_{23}|1\rangle_4|\Omega_{01}\rangle\right)$

$= |0\rangle_1|00\rangle_{23}|0\rangle_4|\Omega_{00}\rangle + |0\rangle_1|11\rangle_{23}|0\rangle_4|\Omega_{00}\rangle + |0\rangle_1|01\rangle_{23}|1\rangle_4|\Omega_{01}\rangle - |0\rangle_1|10\rangle_{23}|1\rangle_4|\Omega_{01}\rangle$

$= \dfrac{1}{\sqrt{2}}\Big[|0\rangle_1|0\rangle_2\left(|\phi^+\rangle_{34} + |\phi^-\rangle_{34}\right)|\Omega_{00}\rangle + |0\rangle_1|1\rangle_2\left(|\psi^+\rangle_{34} - |\psi^-\rangle_{34}\right)|\Omega_{00}\rangle$
$+ |0\rangle_1|0\rangle_2\left(|\phi^+\rangle_{34} - |\phi^-\rangle_{34}\right)|\Omega_{01}\rangle - |0\rangle_1|1\rangle_2\left(|\psi^+\rangle_{34} + |\psi^-\rangle_{34}\right)|\Omega_{01}\rangle\Big]$

$= \dfrac{1}{\sqrt{2}}\Big[|0\rangle_1|0\rangle_2|\phi^+\rangle_{34}\left(|\Omega_{00}\rangle + |\Omega_{01}\rangle\right) + |0\rangle_1|0\rangle_2|\phi^-\rangle_{34}\left(|\Omega_{00}\rangle - |\Omega_{01}\rangle\right)$

$$+|0\rangle_1|1\rangle_2|\psi^+\rangle_{34}(|\Omega_{00}\rangle-|\Omega_{01}\rangle)-|0\rangle_1|1\rangle_2|\psi^-\rangle_{34}(|\Omega_{00}\rangle+|\Omega_{01}\rangle)\big]. \tag{17}$$

For Eve not being detectable in Case (b) of Step 4, Alice's Bell basis measurement result on $S_3^i$ and $S_4^i$, Bob's $Z$ basis measurement result on $S_1^i$ and Alice's $Z$ basis measurement result on $S_2^i$ should obey to Eq.(3); and Alice's $Z$ basis measurement result on the fresh particle corresponding to $S_1^i$ from Bob should be same to the prepared state from Bob. Consequently, it should have

$$|\Omega_{00}\rangle=|\Omega_{01}\rangle. \tag{18}$$

Secondly, consider that Bob's measurement result is $|1\rangle$. After Eve performs $U_F$ on the particles from Bob and Charlie back to Alice, according to Eqs.(9,12,13,16), the state of the composite system is turned into

$$U_F\big(|1\rangle_1|00\rangle_{23}|0\rangle_4|E_{1000}\rangle+|1\rangle_1|00\rangle_{23}|1\rangle_4|E_{1001}\rangle+|1\rangle_1|10\rangle_{23}|0\rangle_4|E_{1100}\rangle+|1\rangle_1|10\rangle_{23}|1\rangle_4|E_{1101}\rangle$$

$$+|1\rangle_1|01\rangle_{23}|0\rangle_4|E_{1010}\rangle+|1\rangle_1|01\rangle_{23}|1\rangle_4|E_{1011}\rangle+|1\rangle_1|11\rangle_{23}|0\rangle_4|E_{1110}\rangle+|1\rangle_1|11\rangle_{23}|1\rangle_4|E_{1111}\rangle\big)$$

$$=U_F\big(|1\rangle_1|00\rangle_{23}|1\rangle_4|E_{1001}\rangle+|1\rangle_1|10\rangle_{23}|0\rangle_4|E_{1100}\rangle+|1\rangle_1|01\rangle_{23}|0\rangle_4|E_{1010}\rangle+|1\rangle_1|11\rangle_{23}|1\rangle_4|E_{1111}\rangle\big)$$

$$=\sqrt{2}U_F\big(|1\rangle_1|\phi^-\rangle_{23}|1\rangle_4|E_{1001}\rangle+|1\rangle_1|\psi^+\rangle_{23}|0\rangle_4|E_{1100}\rangle\big)$$

$$=\sqrt{2}U_F\big(|1\rangle_1|\mathcal{G}_{11}\rangle_{23}|1\rangle_4|\Delta_{11}\rangle+|1\rangle_1|\mathcal{G}_{10}\rangle_{23}|0\rangle_4|\Delta_{10}\rangle\big)$$

$$=\sqrt{2}\big(|1\rangle_1|\mathcal{G}_{11}\rangle_{23}|1\rangle_4|\Omega_{11}\rangle+|1\rangle_1|\mathcal{G}_{10}\rangle_{23}|0\rangle_4|\Omega_{10}\rangle\big)$$

$$=|1\rangle_1|00\rangle_{23}|1\rangle_4|\Omega_{11}\rangle-|1\rangle_1|11\rangle_{23}|1\rangle_4|\Omega_{11}\rangle+|1\rangle_1|01\rangle_{23}|0\rangle_4|\Omega_{10}\rangle+|1\rangle_1|10\rangle_{23}|0\rangle_4|\Omega_{10}\rangle$$

$$=\frac{1}{\sqrt{2}}\Big[|1\rangle_1|0\rangle_2\big(|\psi^+\rangle_{34}+|\psi^-\rangle_{34}\big)|\Omega_{11}\rangle-|1\rangle_1|1\rangle_2\big(|\phi^+\rangle_{34}-|\phi^-\rangle_{34}\big)|\Omega_{11}\rangle$$

$$+|1\rangle_1|0\rangle_2\big(|\psi^+\rangle_{34}-|\psi^-\rangle_{34}\big)|\Omega_{10}\rangle+|1\rangle_1|1\rangle_2\big(|\phi^+\rangle_{34}+|\phi^-\rangle_{34}\big)|\Omega_{10}\rangle\Big]$$

$$=\frac{1}{\sqrt{2}}\Big[|1\rangle_1|0\rangle_2|\psi^+\rangle_{34}(|\Omega_{11}\rangle+|\Omega_{10}\rangle)+|1\rangle_1|0\rangle_2|\psi^-\rangle_{34}(|\Omega_{11}\rangle-|\Omega_{10}\rangle)$$

$$+|1\rangle_1|1\rangle_2|\phi^+\rangle_{34}(|\Omega_{10}\rangle-|\Omega_{11}\rangle)+|1\rangle_1|1\rangle_2|\phi^-\rangle_{34}(|\Omega_{10}\rangle+|\Omega_{11}\rangle)\Big]. \tag{19}$$

For Eve not being detectable in Case (b) of Step 4, Alice's Bell basis measurement result on $S_3^i$ and $S_4^i$, Bob's $Z$ basis measurement result on $S_1^i$ and Alice's $Z$ basis measurement result on $S_2^i$ should obey to Eq.(3); and Alice's $Z$ basis measurement result on the fresh particle corresponding to $S_1^i$ from Bob should be same to the prepared state from Bob. Consequently, it should have

$$|\Omega_{10}\rangle=|\Omega_{11}\rangle. \tag{20}$$

(iii) Assume that Charlie has chosen to SIFT and Bob has chosen to CRTL. Thus, when Charlie's measurement result is $|0\rangle$, the state of the composite system in Eq.(8) is turned into

$|0\rangle_1|00\rangle_{23}|0\rangle_4|E_{0000}\rangle+|1\rangle_1|00\rangle_{23}|0\rangle_4|E_{1000}\rangle+|0\rangle_1|10\rangle_{23}|0\rangle_4|E_{0100}\rangle+|1\rangle_1|10\rangle_{23}|0\rangle_4|E_{1100}\rangle$

$+|0\rangle_1|01\rangle_{23}|0\rangle_4|E_{0010}\rangle+|1\rangle_1|01\rangle_{23}|0\rangle_4|E_{1010}\rangle+|0\rangle_1|11\rangle_{23}|0\rangle_4|E_{0110}\rangle+|1\rangle_1|11\rangle_{23}|0\rangle_4|E_{1110}\rangle$ ; when Charlie's measurement result is $|1\rangle$, the state of the composite system in Eq.(8) is turned into

$|0\rangle_1|00\rangle_{23}|1\rangle_4|E_{0001}\rangle+|1\rangle_1|00\rangle_{23}|1\rangle_4|E_{1001}\rangle+|0\rangle_1|10\rangle_{23}|1\rangle_4|E_{0101}\rangle+|1\rangle_1|10\rangle_{23}|1\rangle_4|E_{1101}\rangle$

$+|0\rangle_1|01\rangle_{23}|1\rangle_4|E_{0011}\rangle+|1\rangle_1|01\rangle_{23}|1\rangle_4|E_{1011}\rangle+|0\rangle_1|11\rangle_{23}|1\rangle_4|E_{0111}\rangle+|1\rangle_1|11\rangle_{23}|1\rangle_4|E_{1111}\rangle$.

Firstly, consider that Charlie's measurement result is $|0\rangle$. After Eve performs $U_F$ on the particles from Bob and Charlie back to Alice, according to Eqs.(9,10,12,16), the state of the composite system is changed into

$U_F\big(|0\rangle_1|00\rangle_{23}|0\rangle_4|E_{0000}\rangle+|1\rangle_1|00\rangle_{23}|0\rangle_4|E_{1000}\rangle+|0\rangle_1|10\rangle_{23}|0\rangle_4|E_{0100}\rangle+|1\rangle_1|10\rangle_{23}|0\rangle_4|E_{1100}\rangle$

$+|0\rangle_1|01\rangle_{23}|0\rangle_4|E_{0010}\rangle+|1\rangle_1|01\rangle_{23}|0\rangle_4|E_{1010}\rangle+|0\rangle_1|11\rangle_{23}|0\rangle_4|E_{0110}\rangle+|1\rangle_1|11\rangle_{23}|0\rangle_4|E_{1110}\rangle\big)$

$=U_F\big(|0\rangle_1|00\rangle_{23}|0\rangle_4|E_{0000}\rangle+|1\rangle_1|10\rangle_{23}|0\rangle_4|E_{1100}\rangle+|1\rangle_1|01\rangle_{23}|0\rangle_4|E_{1010}\rangle+|0\rangle_1|11\rangle_{23}|0\rangle_4|E_{0110}\rangle\big)$

$=\sqrt{2}U_F\big(|0\rangle_1|\phi^+\rangle_{23}|0\rangle_4|E_{0000}\rangle+|1\rangle_1|\psi^+\rangle_{23}|0\rangle_4|E_{1100}\rangle\big)$

$=\sqrt{2}U_F\big(|0\rangle_1|\vartheta_{00}\rangle_{23}|0\rangle_4|\Delta_{00}\rangle+|1\rangle_1|\vartheta_{10}\rangle_{23}|0\rangle_4|\Delta_{10}\rangle\big)$

$=\sqrt{2}\big(|0\rangle_1|\vartheta_{00}\rangle_{23}|0\rangle_4|\Omega_{00}\rangle+|1\rangle_1|\vartheta_{10}\rangle_{23}|0\rangle_4|\Omega_{10}\rangle\big)$

$=|0\rangle_1|00\rangle_{23}|0\rangle_4|\Omega_{00}\rangle+|0\rangle_1|11\rangle_{23}|0\rangle_4|\Omega_{00}\rangle+|1\rangle_1|01\rangle_{23}|0\rangle_4|\Omega_{10}\rangle+|1\rangle_1|10\rangle_{23}|0\rangle_4|\Omega_{10}\rangle$

$=\dfrac{1}{\sqrt{2}}\Big[\big(|\phi^+\rangle_{12}+|\phi^-\rangle_{12}\big)|0\rangle_3|0\rangle_4|\Omega_{00}\rangle+\big(|\psi^+\rangle_{12}+|\psi^-\rangle_{12}\big)|1\rangle_3|0\rangle_4|\Omega_{00}\rangle$

$+\big(|\psi^+\rangle_{12}-|\psi^-\rangle_{12}\big)|1\rangle_3|0\rangle_4|\Omega_{10}\rangle+\big(|\phi^+\rangle_{12}-|\phi^-\rangle_{12}\big)|0\rangle_3|0\rangle_4|\Omega_{10}\rangle\Big]$

$=\dfrac{1}{\sqrt{2}}\Big[|\phi^+\rangle_{12}|0\rangle_3|0\rangle_4\big(|\Omega_{00}\rangle+|\Omega_{10}\rangle\big)+|\phi^-\rangle_{12}|0\rangle_3|0\rangle_4\big(|\Omega_{00}\rangle-|\Omega_{10}\rangle\big)$

$+|\psi^+\rangle_{12}|1\rangle_3|0\rangle_4\big(|\Omega_{00}\rangle+|\Omega_{10}\rangle\big)+|\psi^-\rangle_{12}|1\rangle_3|0\rangle_4\big(|\Omega_{00}\rangle-|\Omega_{10}\rangle\big)\Big].$ (21)

For Eve not being detectable in Case (c) of Step 4, Alice's Bell basis measurement result on $S_1^i$ and $S_2^i$, Alice's Z basis measurement result on $S_3^i$ and Charlie's Z basis measurement result on $S_4^i$ should obey to Eq.(2); and Alice's Z basis measurement result on the fresh particle corresponding to $S_4^i$ from Charlie should be same to the prepared state from Charlie. As a result, it should have

$$|\Omega_{00}\rangle=|\Omega_{10}\rangle.$$ (22)

Secondly, consider that Charlie's measurement result is $|1\rangle$. After Eve performs $U_F$ on the particles from Bob and Charlie back to Alice, according to Eqs.(9,11,13,16), the state of the composite system is changed into

$U_F\big(|0\rangle_1|00\rangle_{23}|1\rangle_4|E_{0001}\rangle+|1\rangle_1|00\rangle_{23}|1\rangle_4|E_{1001}\rangle+|0\rangle_1|10\rangle_{23}|1\rangle_4|E_{0101}\rangle+|1\rangle_1|10\rangle_{23}|1\rangle_4|E_{1101}\rangle$

$$+ |0\rangle_1 |01\rangle_{23} |1\rangle_4 |E_{0011}\rangle + |1\rangle_1 |01\rangle_{23} |1\rangle_4 |E_{1011}\rangle + |0\rangle_1 |11\rangle_{23} |1\rangle_4 |E_{0111}\rangle + |1\rangle_1 |11\rangle_{23} |1\rangle_4 |E_{1111}\rangle )$$

$$= U_F (|1\rangle_1 |00\rangle_{23} |1\rangle_4 |E_{1001}\rangle + |0\rangle_1 |10\rangle_{23} |1\rangle_4 |E_{0101}\rangle + |0\rangle_1 |01\rangle_{23} |1\rangle_4 |E_{0011}\rangle + |1\rangle_1 |11\rangle_{23} |1\rangle_4 |E_{1111}\rangle )$$

$$= \sqrt{2} U_F ( |1\rangle_1 |\phi^-\rangle_{23} |1\rangle_4 |E_{1001}\rangle + |0\rangle_1 |\psi^-\rangle_{23} |1\rangle_4 |E_{0011}\rangle )$$

$$= \sqrt{2} U_F ( |1\rangle_1 |\vartheta_{11}\rangle_{23} 1_4 |\Delta_{11}\rangle + |0\rangle_1 |\vartheta_{01}\rangle_{23} |1\rangle_4 |\Delta_{01}\rangle )$$

$$= \sqrt{2} ( |1\rangle_1 |\vartheta_{11}\rangle_{23} |1\rangle_4 |\Omega_{11}\rangle + |0\rangle_1 |\vartheta_{01}\rangle_{23} |1\rangle_4 |\Omega_{01}\rangle )$$

$$= |1\rangle_1 |00\rangle_{23} |1\rangle_4 |\Omega_{11}\rangle - |1\rangle_1 |11\rangle_{23} |1\rangle_4 |\Omega_{11}\rangle + |0\rangle_1 |01\rangle_{23} |1\rangle_4 |\Omega_{01}\rangle - |0\rangle_1 |10\rangle_{23} |1\rangle_4 |\Omega_{01}\rangle$$

$$= \frac{1}{\sqrt{2}} \Big[ ( |\psi^+\rangle_{12} - |\psi^-\rangle_{12} ) |0\rangle_3 |1\rangle_4 |\Omega_{11}\rangle - ( |\phi^+\rangle_{12} - |\phi^-\rangle_{12} ) |1\rangle_3 |1\rangle_4 |\Omega_{11}\rangle$$

$$+ ( |\phi^+\rangle_{12} + |\phi^-\rangle_{12} ) |1\rangle_3 |1\rangle_4 |\Omega_{01}\rangle - ( |\psi^+\rangle_{12} + |\psi^-\rangle_{12} ) |0\rangle_3 |1\rangle_4 |\Omega_{01}\rangle \Big]$$

$$= \frac{1}{\sqrt{2}} \Big[ |\psi^+\rangle_{12} |0\rangle_3 |1\rangle_4 ( |\Omega_{11}\rangle - |\Omega_{01}\rangle ) - |\psi^-\rangle_{12} |0\rangle_3 |1\rangle_4 ( |\Omega_{11}\rangle + |\Omega_{01}\rangle )$$

$$- |\phi^+\rangle_{12} |1\rangle_3 |1\rangle_4 ( |\Omega_{11}\rangle - |\Omega_{01}\rangle ) + |\phi^-\rangle_{12} |1\rangle_3 |1\rangle_4 ( |\Omega_{11}\rangle + |\Omega_{01}\rangle ) \Big]. \tag{23}$$

For Eve not being detectable in Case (c) of Step 4, Alice's Bell basis measurement result on $S_1^i$ and $S_2^i$, Alice's Z basis measurement result on $S_3^i$ and Charlie's Z basis measurement result on $S_4^i$ should obey to Eq.(2); and Alice's Z basis measurement result on the fresh particle corresponding to $S_4^i$ from Charlie should be same to the prepared state from Charlie. As a result, it should have

$$|\Omega_{11}\rangle = |\Omega_{01}\rangle . \tag{24}$$

By virtue of Eqs.(18,20,22,24), it can be obtained that

$$|\Omega_{00}\rangle = |\Omega_{01}\rangle = |\Omega_{10}\rangle = |\Omega_{11}\rangle = |\Omega\rangle . \tag{25}$$

(iv) Assume that both Bob and Charlie have chosen to CTRL. After Eve performs $U_F$ on the particles from Bob and Charlie back to Alice, according to Eqs.(15,16), the state of the composite system is evolved into

$$U_F \Big[ U_E (|\chi^{00}\rangle_{1234} \otimes |\xi\rangle_E ) \Big] = \frac{1}{2} U_F ( |0\rangle_1 |\vartheta_{00}\rangle_{23} |0\rangle_4 |\Delta_{00}\rangle + |1\rangle_1 |\vartheta_{11}\rangle_{23} |1\rangle_4 |\Delta_{11}\rangle$$

$$+ |0\rangle_1 |\vartheta_{01}\rangle_{23} |1\rangle_4 |\Delta_{01}\rangle + |1\rangle_1 |\vartheta_{10}\rangle_{23} |0\rangle_4 |\Delta_{10}\rangle )$$

$$= \frac{1}{2} ( |0\rangle_1 |\vartheta_{00}\rangle_{23} |0\rangle_4 |\Omega_{00}\rangle + |1\rangle_1 |\vartheta_{11}\rangle_{23} |1\rangle_4 |\Omega_{11}\rangle$$

$$+ |0\rangle_1 |\vartheta_{01}\rangle_{23} |1\rangle_4 |\Omega_{01}\rangle + |1\rangle_1 |\vartheta_{10}\rangle_{23} |0\rangle_4 |\Omega_{10}\rangle ). \tag{26}$$

Inserting Eq.(25) into Eq.(26) produces

$$U_F \Big[ U_E (|\chi^{00}\rangle_{1234} \otimes |\xi\rangle_E ) \Big] = \frac{1}{2} ( |0\rangle_1 |\vartheta_{00}\rangle_{23} |0\rangle_4 + |1\rangle_1 |\vartheta_{11}\rangle_{23} |1\rangle_4 + |0\rangle_1 |\vartheta_{01}\rangle_{23} |1\rangle_4 + |1\rangle_1 |\vartheta_{10}\rangle_{23} |0\rangle_4 ) |\Omega\rangle$$

$$= |\chi^{00}\rangle_{1234} |\Omega\rangle . \tag{27}$$

For Eve not being detectable in Case (d) of Step 4, Alice's *FMB* basis measurement result on $S_1^i$, $S_2^i$, $S_3^i$ and $S_4^i$ should always be $\left|\chi^{00}\right\rangle$. According to Eq.(27), this point is automatically satisfied.

(v) Inserting Eq.(25) into Eq.(16) produces

$$U_F\left(|x\rangle_1 |\vartheta_{xy}\rangle_{23} |y\rangle_4 |\Delta_{xy}\rangle\right) = |x\rangle_1 |\vartheta_{xy}\rangle_{23} |y\rangle_4 |\Omega\rangle, \quad x, y \in \{0,1\}. \tag{28}$$

Inserting Eq.(25) into Eq.(17) produces

$$U_F\left(|0\rangle_1 |00\rangle_{23} |0\rangle_4 |E_{0000}\rangle + |0\rangle_1 |00\rangle_{23} |1\rangle_4 |E_{0001}\rangle + |0\rangle_1 |10\rangle_{23} |0\rangle_4 |E_{0100}\rangle + |0\rangle_1 |10\rangle_{23} |1\rangle_4 |E_{0101}\rangle\right.$$
$$\left. + |0\rangle_1 |01\rangle_{23} |0\rangle_4 |E_{0010}\rangle + |0\rangle_1 |01\rangle_{23} |1\rangle_4 |E_{0011}\rangle + |0\rangle_1 |11\rangle_{23} |0\rangle_4 |E_{0110}\rangle + |0\rangle_1 |11\rangle_{23} |1\rangle_4 |E_{0111}\rangle\right)$$
$$= \sqrt{2}\left(|0\rangle_1 |0\rangle_2 |\phi^+\rangle_{34} - |0\rangle_1 |1\rangle_2 |\psi^-\rangle_{34}\right)|\Omega\rangle. \tag{29}$$

Inserting Eq.(25) into Eq.(19) produces

$$U_F\left(|1\rangle_1 |00\rangle_{23} |0\rangle_4 |E_{1000}\rangle + |1\rangle_1 |00\rangle_{23} |1\rangle_4 |E_{1001}\rangle + |1\rangle_1 |10\rangle_{23} |0\rangle_4 |E_{1100}\rangle + |1\rangle_1 |10\rangle_{23} |1\rangle_4 |E_{1101}\rangle\right.$$
$$\left. + |1\rangle_1 |01\rangle_{23} |0\rangle_4 |E_{1010}\rangle + |1\rangle_1 |01\rangle_{23} |1\rangle_4 |E_{1011}\rangle + |1\rangle_1 |11\rangle_{23} |0\rangle_4 |E_{1110}\rangle + |1\rangle_1 |11\rangle_{23} |1\rangle_4 |E_{1111}\rangle\right)$$
$$= \sqrt{2}\left(|1\rangle_1 |0\rangle_2 |\psi^+\rangle_{34} + |1\rangle_1 |1\rangle_2 |\phi^-\rangle_{34}\right)|\Omega\rangle. \tag{30}$$

Inserting Eq.(25) into Eq.(21) produces

$$U_F\left(|0\rangle_1 |00\rangle_{23} |0\rangle_4 |E_{0000}\rangle + |1\rangle_1 |00\rangle_{23} |0\rangle_4 |E_{1000}\rangle + |0\rangle_1 |10\rangle_{23} |0\rangle_4 |E_{0100}\rangle + |1\rangle_1 |10\rangle_{23} |0\rangle_4 |E_{1100}\rangle\right.$$
$$\left. + |0\rangle_1 |01\rangle_{23} |0\rangle_4 |E_{0010}\rangle + |1\rangle_1 |01\rangle_{23} |0\rangle_4 |E_{1010}\rangle + |0\rangle_1 |11\rangle_{23} |0\rangle_4 |E_{0110}\rangle + |1\rangle_1 |11\rangle_{23} |0\rangle_4 |E_{1110}\rangle\right)$$
$$= \sqrt{2}\left(|\phi^+\rangle_{12} |0\rangle_3 |0\rangle_4 + |\psi^+\rangle_{12} |1\rangle_3 |0\rangle_4\right)|\Omega\rangle. \tag{31}$$

Inserting Eq.(25) into Eq.(23) produces

$$U_F\left(|0\rangle_1 |00\rangle_{23} |1\rangle_4 |E_{0001}\rangle + |1\rangle_1 |00\rangle_{23} |1\rangle_4 |E_{1001}\rangle + |0\rangle_1 |10\rangle_{23} |1\rangle_4 |E_{0101}\rangle + |1\rangle_1 |10\rangle_{23} |1\rangle_4 |E_{1101}\rangle\right.$$
$$\left. + |0\rangle_1 |01\rangle_{23} |1\rangle_4 |E_{0011}\rangle + |1\rangle_1 |01\rangle_{23} |1\rangle_4 |E_{1011}\rangle + |0\rangle_1 |11\rangle_{23} |1\rangle_4 |E_{0111}\rangle + |1\rangle_1 |11\rangle_{23} |1\rangle_4 |E_{1111}\rangle\right)$$
$$= \sqrt{2}\left(|\phi^-\rangle_{12} |1\rangle_3 |1\rangle_4 - |\psi^-\rangle_{12} |0\rangle_3 |1\rangle_4\right)|\Omega\rangle. \tag{32}$$

According to Eqs.(27-32), it can be concluded that for incurring no error in Step 4, the final state of Eve's probe should be independent from Alice, Bob and Charlie's operations and measurement results. Therefore, Eve gets no information on Alice's shared secret key bit. Thus, we have completely proved Theorem 1.

**3.2 The participant attack**

For QSS, it has been indicated in Ref.[75] that if the eavesdropping attack behaviors from internal participants can be discovered, the eavesdropping attack behavior from any eavesdropper (whether internal participants or an external eavesdropper) will automatically be detected. In our protocol, the role of Bob is the same as that of Charlie. Without loss of generality, assume that Bob is an internal attacker with infinite quantum power and wants to get Alice's secret bits without Charlie's help.

(1) Trojan horse attack

The particles in $S_4$ are transmitted from Alice to Charlie, while the fresh particles corresponding to the ones in $S_4$ are sent from Charlie back to Alice. As a result, Bob may adopt the Trojan horse attack, containing the invisible photon eavesdropping attack [76] and the delay-photon Trojan horse attack [77,78], to get something useful. By virtue of Refs.[78,79], Charlie can use a wavelength filter and a photon number splitter to prevent the invisible photon eavesdropping attack and the delay-photon Trojan horse attack from Bob, respectively.

(2) Entangle-measure attack

When Bob launches the entangle-measure attack depicted in Fig.1, the following Lemma can be easily derived from Theorem 1. Lemma 1 implies that although Bob knows his own choice of operation and measurement result, he will still have no access to Charlie's operation and measurement result if he wants to leave no trace of his attack; as a result, Bob cannot know Alice's shared key bit alone.

**Lemma 1:** When Bob performs attack $(U_E, U_F)$ on the particles from Alice to Bob and Charlie and back to Alice, for inducing no error in Step 4, the final state of Bob's probe should be independent of Charlie's operation and measurement result. Therefore, Bob cannot get Alice's shared secret key bit alone.

## 4  Discussions and conclusions

Now we calculate the qubit efficiency defined as [4]

$$\eta = \frac{\lambda_s}{\lambda_q + \lambda_c}, \tag{33}$$

where $\lambda_s$, $\lambda_q$ and $\lambda_c$ are the number of shared classical key bits, the number of consumed qubits and the number of classical bits used for the classical communication, respectively. The classical bits used for security check processes are not taken into account here.

In the proposed SQSS protocol, Alice can successfully share her $n$ secret classical bits with Bob and Charlie, hence $\lambda_s = n$. Alice needs to generate $8n$ initial χ-type states and transmit all of the first (fourth) particles to Bob (Charlie); and when Bob (Charlie) chooses to SIFT, he (she) needs to generate $4n$ fresh qubits and send them to Alice, hence $\lambda_q = 8n \times 4 + 4n \times 2 = 40n$. There are no classical bits used for the classical communication, hence $\lambda_c = 0$. It can be concluded that the qubit efficiency of the proposed SQSS protocol is $\eta = \frac{n}{40n} = \frac{1}{40}$.

In the following, we compare the proposed SQSS protocol with the SQSS protocol in Ref.[74] which also adopts four-particle entangled states as the initial quantum resource. The comparison results are summarized in Table 3.

It can be concluded from Table 3 that on one hand, the proposed SQSS protocol exceeds the SQSS protocol of Ref.[74] on the aspect of initial quantum resource, since it needn't use Bell state as the initial quantum resource; and on the other hand, the proposed SQSS protocol takes advantage over the SQSS protocol of Ref.[74] on the usage of quantum entanglement swapping, since it does not need quantum entanglement swapping.

The proposed protocol utilizes χ-type states to generate the private random string $r_b$ shared between Alice and Bob and the private random string $r_c$ shared between Alice and Charlie, where $r_b = [r_b^1, r_b^2, \ldots, r_b^n]$ and $r_c = [r_c^1, r_c^2, \ldots, r_c^n]$. Here, $r_b$ is kept secret from Charlie, while $r_c$ is kept secret from Bob. Then, Alice encodes her shared secret key as $r_a = r_b \oplus r_c = [r_b^1 \oplus r_c^1, r_b^2 \oplus r_c^2, \ldots, r_b^n \oplus r_c^n]$. As a result, only when Bob and Charlie cooperate together can they recover $r_a$. Apparently, the proposed protocol can check the security of quantum channel between Alice and Bob and the security of quantum channel between Alice and Charlie together, by utilizing the entanglement correlation of different particles in one χ-type state. However, if $r_b$ and $r_c$ are generated by using two separate SQKD protocols, respectively, the corresponding modified protocol has to check the security of quantum channel between Alice and Bob and the security of quantum channel between Alice and Charlie separately. The quantum resource and the classical resource consumed for security checks in the proposed protocol may be less than those of the modified protocol. In addition, there are so many different SQKD protocols at present, such as those of Refs.[31-44], so it is hard to say which one is better between the modified protocol and the proposed protocol, considering the performances on the qubit efficiency, the initial quantum resource, the consumed classical resource, the usage of quantum entanglement swapping, the usage of unitary operations, the quantum measurement of classical parties, the quantum measurement of quantum party, *etc*.

To sum up, in this paper, we put forward an SQSS protocol with χ-type states, where quantum Alice's secret key bits can be successfully recovered only when classical Bob and classical Charlie collaborate together. The proposed SQSS protocol has been proved in detail to be completely robust against an eavesdropper. The proposed SQSS protocol only uses one kind of quantum entangled state as the initial quantum resource, needn't share any private key among different participants in advance, and require neither quantum entanglement swapping nor unitary operations.

Table 3  Comparison results of the proposed SQSS protocol and the SQSS protocol in Ref.[74]

|  | The SQSS protocol of Ref.[74] | The proposed SQSS protocol |
| --- | --- | --- |
| Initial quantum resource | Four-particle Cluster state and Bell state | χ-type state |
| Usage of quantum entanglement swapping | Yes | No |
| Usage of unitary operations | No | No |
| Quantum measurement of classical parties | $Z$ basis measurement | $Z$ basis measurement |
| Quantum measurement of quantum party | $Z$ basis measurement and Bell basis measurement | $Z$ basis measurement, Bell basis measurement and $FMB$ basis measurement |

**Acknowledgments**

Funding by the National Natural Science Foundation of China (Grant No.62071430 and No.61871347) and the Fundamental Research Funds for the Provincial Universities of Zhejiang (Grant No.JRK21002) is gratefully acknowledged.